\documentclass[doublecol]{epl2}
\usepackage{amsmath}
\usepackage{amssymb}
\usepackage{mathtools}
\usepackage{txfonts}
\UseRawInputEncoding

\title{Measurement of the Unusual Dielectric Response to Low-Frequency  s-Polarized
Evanescent Waves in Metals with {\break} Implications for the Casimir Effect}

\shorttitle{Measurement of the Unusual Dielectric Response to Low-Frequency  s-Polarized
Evanescent Waves in Metals}

\author{
 M.~Dhital\inst{1} \and G.~L.~Klimchitskaya\inst{2,3} \and V.~M.~Mostepanenko\inst{2,3}
 \and  U.~Mohideen\inst{1}}

\shortauthor{ M.~Dhital \etal}

\institute{
\inst{1}Department of Physics and Astronomy, University of California, Riverside, California 92521, USA\\
\inst{2}Central Astronomical Observatory at Pulkovo of the
Russian Academy of Sciences, Saint Petersburg,
196140, Russia\\
\inst{3}Peter the Great Saint Petersburg
Polytechnic University, Saint Petersburg, 195251, Russia}

\abstract{
We report precision measurements of the lateral component of the oscillating magnetic
field reflected from a copper plate, which is fully determined by s-polarized
evanescent waves. The measurement data are compared with theoretical predictions
of classical electrodynamics using the dielectric permittivity of copper as given by
the Drude model. It is shown that these predictions are excluded by the measurement
data which means that the currently used Drude model does not  provide a complete
description of the electromagnetic response of metals for
s-polarized evanescent waves. The critical importance of this result for several
fields of condensed matter physics and optics dealing with evanescent waves,
including the Casimir effect, is discussed.
}

\begin{document} \maketitle

\section{Introduction}
Paul Drude proposed his model of electrical conduction of metals
125 years ago. After the necessary modifications taking into account quantum
theory, which were made in 1933 by Arnold Sommerfeld, this model was
extensively used for the description of an immense number of electrical
phenomena in condensed matter physics, optics, and applied physics \cite{1}. According
to the Drude model, the frequency-dependent dielectric permittivity describing
the response of metals to an alternating electromagnetic field is given by
\begin{equation}
\varepsilon(\omega)=1-\frac{\omega_p^2}{\omega[\omega+i\gamma(T)]},
\label{eq1}
\end{equation}
\noindent
where $\omega_p$ is the plasma frequency and $\gamma$ is the
temperature-dependent relaxation parameter which is inverse of the mean free time
for electron-ion collisions.

The dielectric permittivity (\ref{eq1}) has an abundance of experimental
confirmations in both classical and quantum physics as well as in
electrical engineering. In 2003, it was, however, shown that the values of the
Casimir force \cite{2} between metallic surfaces calculated by the fundamental
Lifshitz theory \cite{3,4,5} using the permittivity (\ref{eq1}) at low frequencies
are excluded by the measurement data \cite{6}. Currently the Casimir force
finds extensive applications in both fundamental
physics and nanotechnology (see the monographs \cite{7,8,9,10,11} and
reviews \cite{12,13,14,15,15a}. Like the van der Waals force, it is determined by
quantum fluctuations of the electromagnetic field, but acts at larger
separations between the interacting bodies, where the effects of relativistic
retardation contribute to the result.

The experiment \cite{6} was followed by experiments of increasing precision
performed by means of a micromechanical torsional oscillator \cite{16,17,18,19,20}
and by an atomic force microscope \cite{21,22,23,24,25,26,27}.
{These experiments used different preparation details of thin metallic (Au) layers
and different measurement techniques, but no discrepancy in the obtained
data was observed.}
In all these
experiments, the predictions of the Lifshitz theory obtained using the permittivity
of metals  (\ref{eq1}) were excluded by the measurement data. Thus, in the
differential force measurement \cite{19}, the difference between the theoretical
results and the experimental data was confirmed by up to a factor of 1000.
The limitations of the permittivity  (\ref{eq1}) for a description of the
low-frequency electromagnetic response of Au in calculations of the Casimir
force were confirmed up to separations of 1.1~$\mu$m \cite{27} and
4.8~$\mu$m \cite{20} between the interacting surfaces. It was shown  \cite{27a}
that taking into account the dependence of the relaxation parameter in eq.~(\ref{eq1})
on the frequency (the so-called Gurzhi model) does not bring the theoretical
predictions in the above experiments into agreement with the measurement data.
Only one measurement of the Casimir force at large separations stated
that the data are in agreement with theory using the Drude model \cite{28}, but
this conclusion was reached by omitting the background force of unknown
origin which exceeded the Casimir force by an order of magnitude and disregarding the
role of imperfections on the surface of a glass lens of centimeter-size
radius \cite{29,30}.

According to the Lifshitz formula for the Casimir force written along the real
frequency axis, both the propagating (on-the-mass-shell where for the photon
frequency and momentum equation $\omega=kc$ holds) and evanescent
(off-the-mass-shell where $\omega\neq kc$) waves of both s- and p-polarizations contribute
to the result. These contributions were analyzed by several authors \cite{31,32,33,34,35}.
It is now well established \cite{36} that the difference between theoretical predictions
and the measurement data is completely determined by the contribution of the
s-polarized (transverse electric) evanescent waves.

It should be noted that most of experimental confirmations of the permittivity
(\ref{eq1})  fall in the region of propagating waves. Physics of surface plasmon
polaritons \cite{37,38} provides a lot of data regarding the response of metals
to the evanescent waves, but for only the p-polarized case. The same is true regarding
near-field optical microscopy \cite{39,40,41,42} used to surpass the standard
resolution limit. Regarding the phenomena of total internal reflection and
frustrated total internal reflection \cite{43,44,45}, they allow testing the response
of metals to electromagnetic waves with very small deviations from the
mass-shell equation. This means that in the region of strongly evanescent
s-polarized waves the dielectric permittivity (\ref{eq1}) has not been
experimentally confirmed.

As a potential test, an experiment sensitive to the reflection coefficient for
s-polarization in the range of strongly evanescent waves has been
proposed \cite{46,47} which uses an oscillating magnetic dipole placed at
height $h$ above a thick metallic plate. It was shown that the lateral
component of emitted magnetic field observed at the height $z=h$ at a
 sufficiently large distance $x$ from the dipole is completely determined by
the s-polarized evanescent waves. Thus, it is possible to directly test the model
of the electromagnetic response of a metal to the s-polarized evanescent waves
by comparing the theoretical results with measurement data.

In this letter, we experimentally probe the response of metals to the low-frequency
s-polarized evanescent waves by measuring the magnitude of lateral component of
the magnetic field reflected from a copper surface. According to our results, the
theoretical values of the reflected field component computed at different
separations from the emitter using the dielectric permittivity (\ref{eq1}) do not
provide an accurate description of the measurement data. Possible implications
of this result in the
Casimir effect and, more widely, in several areas of condensed matter physics
and optics dealing with evanescent waves are discussed.

\section{Configuration}
We consider the electromagnetic field emitted by a coil
(magnetic dipole) spaced at the height $z=h$ above the coordinate origin
$x=0, y=0$ on a thick metallic plate characterized by the dielectric permittivity
$\varepsilon(\omega)$. The magnetic moment of this dipole oscillates with
the frequency $\omega_d$ and is directed along the $z$ axis
\begin{equation}
{\bf m}=(0, 0, m_0e^{-i\omega_dt}).
\label{eq2}
\end{equation}
\noindent
It is assumed that the dipole size is much less than $h$, the separation distance
$r$ between the dipole and the observation point of the emitted field, and the
wavelength $\lambda_d=2\pi c/\omega_d$. It is also assumed that $r \ll \lambda_d$.
As a result, the components of the electric field at the observation point are
smaller than those of the magnetic field by the factor of $\lambda_d/(2\pi r)\sim10^9$
\cite{46,47,48}.

Under these conditions, the $x$-component of the magnetic field at a point $(x,y,z)$
separated from the dipole by the distance $r=[x^2+y^2+(z-h)^2]^{1/2}$ is given
by \cite{46,47}
\begin{eqnarray}
&&B_x(\omega_d,r)=\frac{\mu_0m_0x}{4\pi}\left[\frac{1}{\rho}\!\!\int_{\omega_d/c}^{\infty}\!\!\! dk_{\bot}k_{\bot}^2
J_1(k_{\bot}\rho)R_{\rm s}(\omega_d, k_{\bot}) \right.
\nonumber\\
&&~\left.\times\,\,
e^{-q(z+h)}
-\frac{z-h}{r^2}\left(\frac{\omega_d^2}{c^2r}+3i\frac{\omega_d}{cr^2}-\frac{3}{r^3}
\right)\,e^{i\frac{\omega_d}{c}r}\right]\! ,
\label{eq3}
\end{eqnarray}
\noindent
where $\rho=(x^2+y^2)^{1/2}$, $q^2=k_{\bot}^2-\omega_d^2/c^2$, and
$k_{\bot}$ is the magnitude of the wave vector projection
on the plane of metallic plate, $J_1(z)$ is the Bessel function, and $\mu_0$ is the
magnetic permeability of vacuum. The reflection coefficient on
a metallic plate of sufficiently large thickness $D$
for the s-polarized electromagnetic waves is defined as \cite{10}
\begin{equation}
R_{\rm s}(\omega_d, k_{\bot})=
\frac{q^2-p^2}{q^2+p^2+2qp\coth(pD)},
\label{eq4}
\end{equation}
\noindent
where $p^2=k_{\bot}^2-\varepsilon(\omega_d)\omega_d^2/c^2$.
The component $B_y(\omega_d,r)$ of the magnetic field is obtained from eq.~(\ref{eq3})
by the replacement $x \rightarrow y$. Note that all fields here and below depend on $t$ as
${\rm exp}(-i\omega_d t)$.

As is seen in eq.~(\ref{eq3}), the reflected magnetic field from the plate is
fully determined by the evanescent waves. According to the results of \cite{46,47},
the contribution of the propagating waves to the reflected magnetic field is a
factor of $\lambda_d^3/(2\pi r)^3 \sim 10^{27}$ smaller than that of the
evanescent waves.

\section{Experimental setup}
The experimental setup, shown in fig. 1, involves two coils
(magnetic dipoles): one for generating the low frequency magnetic field (emitter) and the
second for measuring the reflected magnetic field from a copper plate (sensor). An AC
current was applied to the first coil, i.e. ``emitter," to produce the oscillating magnetic field.
This field was then reflected by a $D=2.5~$cm thick copper plate (30 x 30 cm$^2$ area), placed
at a distance $h$ mm below the emitter. The second coil, i.e. ``sensor," was placed at a fixed
distance of $x$ mm from the emitter and at the same height $h$ above the copper plate to
detect the reflected magnetic field. The distances were measured from the centers of the coils
to the top surface of the plate. The emitter coil axis and the copper plate surface were oriented
perpendicular to each other, while the sensor coil axis was perpendicular to the emitter axis
and parallel to the copper plate surface.

Most often, the test bodies used in precision measurements of the
Casimir force between metals were made of gold
\cite{16,17,18,19,20,21, 25,26,27}. Gold is
the suitable material primarily because its chemical inactivity which
allows the surface properties to be stable during the experiment. In
addition, gold also has the beneficial property of very high electric
conductivity and associated excellent optical reflectivity. In this
experiment, the copper plate was used as the model material for the
following reasons. First, a material with very high electrical
conductivity is needed. High chemical purity copper (used here) has
30\% higher electrical conductivity than gold. Second, if there are
deviations from the theory using the Drude and plasma models, they would
be most strongly exhibited at low frequencies below 100~Hz, where the
large plate thickness on the few cm scale is needed based on the large
skin depth. In doing so, the area has to be at least
$30\times30~\mbox{cm}^2$ for
the $h$ and $x$ distances needed. A high purity gold plate with such
dimensions would be prohibitively expensive. For these two reasons, the
gold plate was not used in this experiment. In addition one should not
expect big discrepancy in the measurement results obtained for different
good conductors.

The coordinates of the emitter and sensor centers in the experimental configuration are
$(0,0,h)$ and $(x,0,h)$, respectively. Then, according to eq.~(\ref{eq3}), the lateral component
of the magnetic field at the observation point is completely determined by the part reflected
from the plate
\begin{equation}
B_x(\omega_d,x)=\frac{\mu_0m_0}{4\pi}\!\int_{\omega_d/c}^{\infty}\!\!\! dk_{\bot}k_{\bot}^2
J_1(k_{\bot}x)R_{\rm s}(\omega_d, k_{\bot}) e^{-2hq},
\label{eq5}
\end{equation}
\noindent
whereas $B_y=0$.

The first and most crucial step in this experiment is to create the appropriate magnetic dipoles
for the generation of the magnetic field with the emitter coil as well as the measurement of the
 lateral component of the reflected magnetic field with the sensor coil. The challenge in creating
an appropriate magnetic dipole for the magnetic field generation is that it must approximate a
point dipole in comparison to the other dimensions $x$ and $h$ in the setup. This means that the
radius of the coil should be physically much less than $x$ and $h$ while still be capable of generating
 a sufficiently strong magnetic field that can be detected after reflecting off a metal plate placed
at  large $x$ and $h$.  Therefore, the coil needs to be rather small in size while being capable of
handling large currents greater than $I=1$~A to generate the required stronger magnetic fields.
 A second challenge is designing a sensor coil again of very small size similar to the emitter but
 that can detect extremely weak magnetic fields (on the nT scale) reflected from the copper plate at
large distances compared to the size of the coil.

Two different copper coils, similar in physical size but with different wire diameters and numbers
of turns, were used for the emitter and sensor, respectively. The emitter had a smaller effective
area due to the smaller number of turns but could handle larger currents due to the larger wire
diameter which allowed it to produce the strong magnetic fields.  A commercially available copper
coil (DigiKey part N 732-11702-ND) having 85 turns with an outer radius of 2.6 mm, inner radius
of 1.7 mm, and height of 3.3 mm was used as the emitter.  To prevent the overheating and
melting of the emitter for the large currents used it had to be cooled by blowing dry liquid
nitrogen vapor. The liquid nitrogen vapor was generated by running nitrogen gas through
liquid nitrogen contained in an insulated tank.

The sensor had a larger effective area due to the larger number of turns even though the physical
size remained similar (outer radius of 2.7 mm, inner radius of 1.6 mm and height of 5 mm). The
sensor coil was fabricated using 40 AWG copper wire (Remington Inc., MW 79-C) wound on a
ferrite core (DigiKey part N 1934-1347-ND). The wire was wound tightly on a 5 mm long ferrite
 rod, ensuring each turn was tight, circular and planar. Very circular and planar windings of the
 wire are necessary to achieve high sensitivity and good alignment of the sensor coil. The
resulting coil, with 500 turns, allowed detection of nT magnetic fields.
\begin{figure}[t]
\onefigure{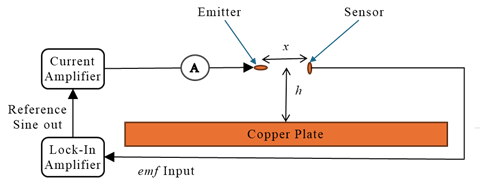}
\caption{
Schematic of the experimental setup. Picture not to scale. The Lock-In Amplifier supplies
the sinusoidal reference signal at the desired frequency as well as measures the amplitude and
phase of the emf generated in the sensor coil from the lateral component of the reflected
magnetic field. The emf measured is calibrated to the magnetic field as discussed in the text.
The current from the Current Amplifier is measured using the ammeter shown as A. The
 emitter was cooled by blowing dry liquid nitrogen vapor to allow the use of large currents.
}
\label{fg1}
\end{figure}

\section{Calibration} Now we discuss the calibration of the emitter and sensor
effective areas, $A_{\rm eff}$, through generation of static magnetic field
and  measurement of the dipole moments of the emitter and sensor
coils
by using a calibrated Gaussmeter (Lakeshore Model 460). A  DC current, $I$, was applied to
the coil, and the generated magnetic field $B(z)$ was measured at different heights $z$ along
the vertical axis of the dipole using the Gaussmeter.
To cancel the background magnetic field, at each value of $z$,
the average of the two measurements with currents running clockwise and counterclockwise
were used. The data at large $z$ were fit to the theoretical axial magnetic field $B(z)$ for the
ideal dipole using the equation \cite{49}
\begin{equation}
B(z)=\frac{\mu_0}{2\pi}\frac{m_0}{z^3}
\label{eq6}
\end{equation}
\noindent
and corresponding magnetic dipole moments $m_0$ of the coils were found. From the
dipole moments, for the current $I$ the effective areas $A_{\rm eff}$ of the coils were
calculated from $A_{\rm eff}=m_0/I$. As a result, $A_{\rm eff}$ of the emitter and sensor
were found to be 1.40 $\times 10^{-3}$ m$^2$ and 2.49 $\times 10^{-2}$ m$^2$, respectively.

\section{Experimental procedure}
Special attention was paid to the background noise and its
reduction. Given the sensitivity of the measurements and the low level of the magnetic fields
being detected, even small amounts of noise could significantly affect the results. To minimize
background noise, all components, including the coils, leads, holders, cables, and connectors,
were properly grounded. The wires connecting the coils were tightly braided  to prevent the
generation or being affected by stray magnetic fields.  Cables connecting the power source
and Lock-In Amplifier were enclosed in stainless steel tubes and securely fixed. The background
noise value without the copper plate was then measured on the Lock-In Amplifier and subtracted
from all measurements.  This value was 21.21 nV  and found to be independent of frequency.

Then the emitter and sensor positioning and alignment were performed. For this purpose,
both the emitter and sensor coils were mounted on the ends of a 28 cm long ceramic tubes
with a diameter of 9 mm.  The ceramic tube was held by a $xyz$ micrometer controlled
positioning stage fixed to the optical table. The coils were positioned at least 30 cm above
the optical table. Care was taken to remove any metallic reflecting objects near the experiment.
First the coils were placed next to each other with their axes aligned and parallel to the plane of
the optical table. Next the current was turned on to the emitter at the experimental frequency $f_d = \omega_d/(2\pi)$.
For the emitter current the reference sine output of the Lock-In Amplifier was amplified using a
EuroPower EP2000 power current amplifier and applied to the emitter coil. An ammeter
 (Agilent 34411A Digital Multimeter) was used to measure the current flowing through the emitter
coil. As noted above, the emitter coil was cooled by blowing dry liquid nitrogen vapor to prevent
its melting.

Then the phase reading on the Lock-In Amplifier was set to zero. Next the axis of the emitter coil
was rotated by 90 degrees to be perpendicular to the plane of the table. By using flat plastic
blocks on top of the optical table, the emitter coil axis was confirmed to be perpendicular to the
optical table. Next small corrections to the sensor coil axis orientation were done to make the
measured emf signal zero.  This confirmed that the axis of the sensor was perpendicular to that
of the emitter.  The sensor coil was moved to the needed $x$ distance from the emitter using
the $x$ axis micrometer. The perpendicular orientation of the sensor axis was further checked
by minor adjustments to its plane by zeroing the sensor emf signal. Then, a 2.5 cm thick,
30 $\times 30$ cm$^2$ copper plate (Alloy 101: Oxygen-Free Electronic, Sequoia Brass and
Copper Inc.) was placed on the optical table at a distance $h$ below the coil centers. The
distance $h$ to the copper plate was changed by using flat insulating plastic blocks on the
optical table below the copper plate. This ensures that the copper plate is parallel to the optical
table and the sensor axis while being perpendicular to the emitter axis. This was also
confirmed by sliding a perpendicular $L$ scale.

The data collection was made as follows. Prior to the experiment, the copper plate was
cleaned with acetic acid, acetone, and methanol
to remove oxide layers and organic contaminants. It was then electrically grounded to the
universal ground. An AC current with a desired frequency and amplitude was supplied to the
 emitter coil. The magnitude and phase angle of the induced emf on the sensor coil, caused by
the lateral component of the reflected magnetic field from the copper plate, were measured
using the Lock-In Amplifier (SR830 DSP). The time constant for signal averaging on the Lock-In
Amplifier was set to 1s. To minimize noise, the Lock-in's inbuilt sync filter was used, and the
dynamic reserve was set to ``low noise." The system was allowed to equilibrate for 2 minutes
before taking the first measurement. A total of 10 measurements were recorded at 15-second
intervals for both magnitude and phase angle of the induced emf.

The copper plate was then moved to a different distance $h$ below the coils while maintaining
the same distance $x$ between the coils. After each height adjustment, the copper plate was
 removed to confirm the emitter and sensor coil orientations were perpendicular to each other
and that there was no additional background noise signal. Once all measurements were taken
for a set of  heights $h$ at a particular current frequency $f_d$, the process was repeated for
other frequencies. Next, the distance between the coils $x$ was changed to another desired
value using the micrometer, and measurements were repeated for different emitter-sensor
heights $h$ above the plate and current frequencies $f_d$.  The experiment was conducted for
inter coil distances $x$ = 42, 46, 50, 54, 58, and 62 mm, and plate-coil heights $h$ = 42, 50,
and 62 mm, at current frequencies $f_d$ = 12, 15, 20, 25, and 50 Hz. The measured emf,
${\cal E}$, was converted to the lateral magnetic field $B _x$ using $B_x = {\cal E}/(2\pi if_d A_{\rm eff})$
\cite{49}, where $A_{\rm eff}$ is the effective area of the sensor coil discussed above and $f_d$ is
the frequency of the signal in Hz.

The mean measured values of  $|B_x|$ are shown by crosses in fig. 2
as the function of separation between the emitter and sensor for $f_d$ equal to (a) 15~Hz, (b) 25~Hz,
and (c) 50~Hz. The vertical arms of the crosses indicate the random errors in the mean lateral
magnetic field which were found at the 67\% confidence level. The systematic errors, which are due
to the error from sensor coil emf measurement, the error from measurement of frequency, and
the error from measurement of the effective area of sensor coil, are much less than
the random error. The horizontal arms of the crosses show the error in measuring separations
with the Vernier calipers, $\Delta x = 0.2$~mm.

\section{Comparison with theory}
The theoretical values of the lateral magnetic field in the
experimental configuration were computed with eq.~(\ref{eq5}) using the dielectric permittivity
of the Drude model (\ref{eq1}) and the reflection coefficient (\ref{eq4}). Note that the presence of
hard plastic blocks under the copper plate does not influence the obtained results.
The value of the plasma
frequency for Cu, $\omega_p = 1.12 \times 10^{16}$~rad/s \cite{50}, has been used. The
value of the relaxation parameter, $\gamma = 1.93 \times 10^{13}$~rad/s, was determined
from the conductivity of the Cu sample, $\sigma = 5.77 \times 10^7$~S/m, indicated by the
manufacturer.
The error in the theoretical values is mostly determined by the errors in $x$ and $h$,
$\Delta x = \Delta h = 0.2$~mm, whereas the error in the value of
$\gamma$ was found to be negligible. The theoretical predictions for the magnitude of
the lateral magnetic field $|B_x|$ are
presented as bands with the widths equal to twice the total theoretical error. To find
the width of the theoretical band, for each set of the parameters, the computation was repeated
with the values $x \pm \Delta x, h \pm \Delta h$, and the smallest and largest of the four
values were selected.

\begin{figure}[t]
\onefigure{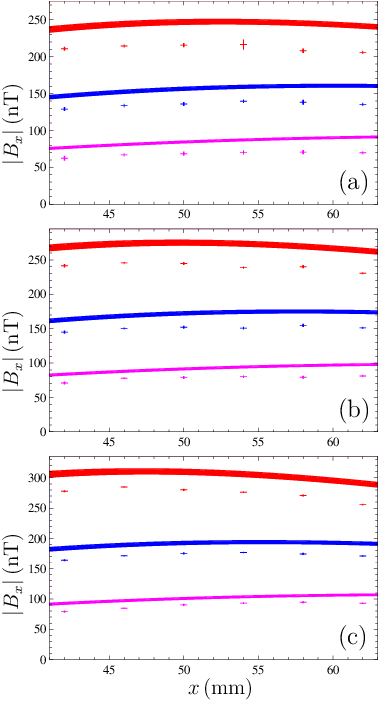}
\caption{
The mean magnitude of the lateral component of oscillating magnetic field reflected from
a copper plate is shown by crosses as a function of separation from the emitter for the
oscillation frequencies (a) 15~Hz, (b) 25~Hz, and (c) 50~Hz. The theoretical bands for the
absolute value of the reflected field are computed at the same frequencies
using the dielectric permittivity
of the Drude model. The bands and the corresponding sets of crosses counted from top
to bottom are plotted for the emitter heights of 42, 50, and 62~mm.
}
\label{fg2}
\end{figure}
\begin{figure}[t]
\onefigure{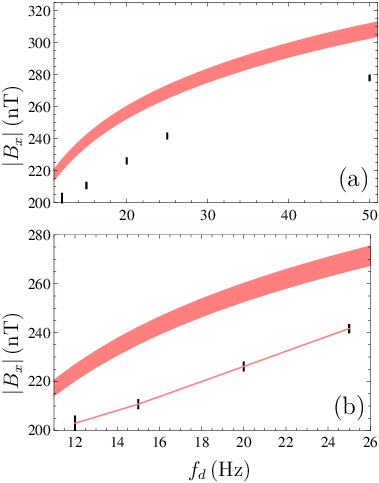}
\caption{
The mean magnitude of the lateral component of oscillating magnetic field reflected
from a copper plate is shown by crosses as a function of  the emitter  frequency
at the height $h=42~$mm and separation from the emitter$x=42~$mm.  The theoretical
band for the absolute value of the reflected field is computed under the same
conditions using the dielectric permittivity of the Drude model within the frequency
regions (a) from 11 to 51~Hz and (b) from 10 to 26~Hz.
}
\label{fg3}
\end{figure}

The theoretical bands obtained for the magnitude of magnetic field reflected from a copper
plate are shown in fig.~2(a,b,c) as a function of separation from the emitter
for different values of the dipole oscillation frequency
(a) $f_d = 15$~Hz, (b) $f_d = 25$~Hz, and (c) $f_d = 50$~Hz. The bands counted from
top to bottom are plotted for the emitter magnetic dipole heights above the copper plate
$h$ = 42, 50, and 62~mm.

In fig.~3(a), the theoretical band for the magnitude of the reflected magnetic field is shown as a function
of the dipole oscillation frequency. The measured values for the magnitude of the reflected magnetic field
are shown by crosses as in fig.~2. For better visualization, the region of frequencies
from 10 to 26~Hz is reproduced on an enlarged scale in fig.~3(b).

From figs.~2 and 3 it can be observed that for all the three frequencies of the emitted magnetic
field, heights above the copper plate, and for all separation distances to the observation
point from 42 to 62~mm the theoretical predictions for the magnitude
of the lateral reflected magnetic field are excluded by the measurement data. Furthermore,  the
dependence of the measurement data on frequency shown in fig. 3(b) is qualitatively different
from that predicted by the theory. What this means is that
the Drude model provides an incomplete representation of the response of metals to
low-frequency s-polarized evanescent waves.

\section{Conclusions and discussion}
The performed experimental test for the completeness
of the Drude model in the area of s-polarized evanescent waves is entirely classical.
By contrast, the Casimir effect at short separations currently tested experimentally is
a quantum phenomenon. The lateral component of magnetic field reflected from
metallic plate is completely determined by the s-polarized evanescent waves,
i.e., by the region where the Drude model lacks experimental confirmation.
In the case of the Casimir force, the contribution of s-polarized evanescent waves
is responsible for the disagreement between the measurement data and theoretical
predictions using the Drude model.
However, it is important to note that for the s-polarized evanescent waves
contribution using the Drude model
the values of the characteristic parameter $ck_{\bot}/\omega$ giving
the dominant fraction of the lateral component of magnetic field and that
responsible for the disagreement in the measurements of the Casimir force
are different. Thus,
at the typical height of $h$ = 5~cm, a 95\% contribution to the lateral magnetic field
coming from the s-polarized evanescent waves is given by the interval
$ck_{\bot}/\omega_d$ from 10$^7$ to 10$^9$.  At the typical separation of 500~nm between
the interacting surfaces, the  95\% of the s-polarized evanescent waves contribution to the
Casimir force comes from the interval $ck_{\bot}/\omega$
from 10$^2$ to 10$^4$. In both cases, the strongly evanescent waves are responsible for
the disagreement between experiment and theory, but there is a five orders of magnitude
 difference in their regions of contribution.

Given the above, it appears that the understanding of the response of metals to
s-polarized evanescent waves is not complete and some modifications in the
permittivity given by the Drude model (\ref{eq1}) might be necessary.
The case of graphene, whose
low-frequency dielectric response is found from the first principles of
quantum field theory \cite{51,52,53,54,55} and confirmed by the experimental data
of the Casimir force measurements in graphene systems \cite{56,57}, suggests that
this dielectric function should be spatially nonlocal and possess a double pole
at zero frequency \cite{58}. This may explain why the Casimir force experiments
\cite{6,16,17,18,19,20,21,22,23,24,25,26,27} are
in agreement with theoretical predictions of the Lifshitz theory if the low-frequency
response of metals is described by the plasma model [i.e., by eq.~(\ref{eq1}) with
$\gamma = 0$], which does not take dissipation into account and should not be
applicable at low frequencies. Mention should be made that the magnitudes of the
reflected lateral magnetic field computed with eq.~(\ref{eq5}) using the plasma
model lie 26\% to 93\% higher depending on the values of $f_d, x$ and $h$ than
the theoretical bands in figs.~2 and 3, i.e., they are in
even stronger disagreement with the measurement data than those found using the
Drude model. This means
that a successful use of the plasma model for calculation of the Casimir force is
merely due to its fortuitous proximity to the true dielectric function in the region
of parameters characteristic for the Casimir effect.

The search for the complete dielectric response  of metals to the s-polarized
evanescent waves in different physical systems is currently under active
investigation (see, for instance, \cite{58a,59,60,61,61a,62,63,64}).
{Future work will enable the
development of the spatially nonlocal permittivity that will fully
describe the dielectric response of metals to both  propagating and
evanescent waves of any polarization. The development of such a
permittivity will be performed starting from the first principles of
quantum electrodynamics along with the measurement data of
many experimental tests including this experiment and experiments on
measuring the Casimir force.}
The resolution
of this problem will impact
research in the areas of nanophotonics, optical quantum computing on a chip,
near-field optical microscopy and its applications
to overcoming the standard resolution limit, physics of total internal
reflection and surface plasmon polaritons, to say nothing of the
Casimir effect and related quantum phenomena of atomic friction and radiation
heat transfer.

\acknowledgments{We are grateful to V.~B.~Svetovoy for useful discussions.
The work of M.D. and U.M. was partially supported by the NSF Grant No PHY-2012201.
The work of G.L.K.\ and V.M.M.~was supported by the State Assignment for Basic Research
(project FSEG-2023-0016)}.

\vspace*{2mm}

{\it Data availability statement:}
All data that support the findings of this study are included within the article.

\end{document}